\documentclass{aa}

\usepackage[varg]{txfonts}
\usepackage{graphicx}
\usepackage{natbib}
\bibpunct{(}{)}{;}{a}{}{,} % to follow the A&A style
\usepackage{amssymb}
\usepackage{amsmath}
\usepackage{multirow}
\usepackage{appendix}
\usepackage[colorlinks=true,     linkcolor=blue, citecolor=blue, filecolor=blue, urlcolor=blue]{hyperref}

\defcitealias{Cherry_2025}{C25}% Define a citation alias

\begin{document}

\title{Decomposing wave activity in the solar atmosphere}
\subtitle{Shocks, jets and swirls in the quiet Sun}

\author{George Cherry\thanks{\email{georgche@uio.no}}\inst{1,2}  \and Boris Gudiksen\inst{1,2} \and Adam J. Finley\inst{3} \and Quentin Noraz\inst{1,2}}

\institute{Rosseland Centre for Solar physics,Universitetet i Oslo, Sem Sælands vei 13, 0371,Oslo, Norway \and Institutt for Teoretisk Astrofysikk, Universitetet i Oslo, Sem Sælands vei 13,0371,Oslo, Norway \and Universit\'e Paris-Saclay, Universit\'e Paris Cit\'e, CEA, CNRS, AIM, 91191, Gif-sur-Yvette, France}

\abstract{There remains much mystery about how or if wave-energy in the photosphere can be transferred sufficiently upwards through the solar atmosphere to contribute to coronal heating. In light of a plethora of theoretical and idealised studies, we must complement our understanding with realistic and self-driven simulations in order to confidently quantify such contributions.}{In this study we aim to connect the various environments from the photosphere to low corona and identify wave drivers, transitions and dissipation mechanisms. We will analyse the effects of the presence of twisted magnetic features and vortical flows on the transport of such wave modes as the structures evolve.}{We adopt the most significant frequency (MSF) decomposition method to trace wave activity through a 3D realistic quiet Sun simulation. We focus on vertical and temporal evolution, identifying wave sources and shifts in the dominant modes.}{We identify two frequencies, at 3.5 and 5~mHz, that connect oscillations in the upper convection zone to the dynamics in the solar atmosphere. We see distinct differences in the absence and presence of swirling structures on the upwards propagation of these oscillations. Furthermore, we validate the use of the highest frequency MSFs as a proxy for the location of shocks in the chromosphere, and use the results to understand the connection between shocks and the propagation of oscillations in the upper atmosphere. We discuss the relation of energy transfer via shocks, mode conversion, and jets. Finally, we find the contribution of 3.5 and 5 mHz signals to the overall wave power in the domain to be significant, up to $50\%$.}{}


\maketitle

%% main text
%%%%%%%%%%%%%%%%%INTRO%%%%%%%%%%%%%%%%%%%%%%%%%%%%%%%%%%%%%%%%

\section{Introduction}\label{sec:intro}
The key to solving the coronal heating problem lies in disambiguating the proposed mechanisms that could deposit energy into the solar corona. The dissipation of magnetoacoustic waves, triggered by convective motions, into heat remains a promising mechanism  \citep{Biermann_1946,Erdelyi2007,Arregui_2015}. MHD theory \citep[e.g.,][]{morton_2012,goossens_2003} contains multiple dissipation processes for waves in inhomogeneous plasma, such as shocks \citep{Park_2019,Snow_2021} and phase mixing \citep{Heyvaerts_1983,Cally_2017,McMurdo_2023}. Energy loss due to thermal conduction, ambipolar diffusion, and viscous heating can also trigger wave dissipation and heating of the surrounding plasma \citep{Khodachenko_2004,Soler_2015,Arregui_2015,Popescu_2021,mikhalyaev_2023}. \\

Whilst waves are powerful energy transfer mechanisms, it is unclear how wave energy is transported across the sharp gradients in the solar atmosphere, such as the equipartition level, where the sound and Alfvén speeds are equivalent, and the transition region, where the conditions for wave propagation change dramatically \citep{morton_2012}. Furthermore, when the plasma-$\beta$ (the ratio of plasma and magnetic pressures) approaches unity, there is no clear dominant restoring force between the plasma pressure and magnetic field making the distinction between wave modes less clear. It is usually assumed that mode conversion in these regions allows the nature of waves to change as they pass through \citep{Bogdan_2003,Newington_2010,Finley_2022}. For example, \citet{Newington_2010} discuss the theoretical and numerical results of mode conversion from gravity modes into magnetically-aligned acoustic modes and Alfvén (magnetic) modes through the equipartition level. This region is often treated analogously with the $\beta = 1$ level, since $\beta=1.2$ at the equipartition level for $\gamma=5/3$.\\

In addition to waves, coronal jets (e.g. spicules) and swirls are thought to play a role in transporting energy in the solar atmosphere. Twisted magnetic features and vortical flows, observed as chromospheric swirls, are capable of transporting photospheric and chromospheric plasma up to coronal heights \citep{Vigeesh_2012,Finley_2022,tziotziou_2023}. Twisted flux tubes can channel magnetoacoustic waves, such as kink and sausage modes along the swirl boundary \citep{Fedun_2011}, as well as producing acoustic waves from the interaction of two vortices \citep{Kitiashvili_2011}, and inducing Alfvén waves \citep{kannan_2024}. %\textcolor{red}{Something about jets/spicules}\\

Simplified simulations have been successful in reproducing and expanding upon MHD theory.  \citet{Bogdan_2003}, for example, demystify the propagation and conversion of MHD waves across the $\beta \approx 1$ level by driving specific waves from the bottom of a 2D stratified domain. \citet{Newington_2010} describe the reflection and conversion of magnetogravity waves  using a 1D chromospheric model, and \citet{riedl_2021} use a cylindrical coordinate system to simulate driven acoustic waves along the length of a 3D coronal loop. However, quantifying wave activity in realistic solar atmosphere simulations is not so simple. Despite several well-established methods for decomposing wave activity within a simulated domain, finding and tracking localised MHD waves remains a complex and challenging task. Our previous study \citet{Cherry_2025}, hereafter \citetalias{Cherry_2025},  introduced the Most Significant Frequency (MSF) decomposition, that can assess the dominant wave frequency in localised regions of a realistic simulation, alongside wave dissipation. We applied the MSF algorithm to a quiet Sun simulation produced with the \verb|Bifrost| stellar atmosphere code. In doing so, we concluded that magneto-acoustic waves can travel along the swirl boundaries, where the plasma-$\beta\approx 1$, and that as the waves diffuse outwards from the boundaries, they are subject to viscous damping, which heats the corona. This study aims to further explore the relationship between waves in the lower solar atmosphere and the corona by applying the MSF approach as a function of height. 

\section{Numerical simulation} \label{sec: Implem}
This study is a continuation of the analysis from \citetalias{Cherry_2025}, which uses a highly-resolved quiet Sun model, produced by the \verb|Bifrost| code. This MHD code aims to realistically simulate the solar atmosphere from the convection zone to the low-corona, including additional terms such as radiative losses and thermal conduction. A summary of the code can be found in \citetalias{Cherry_2025}, and an extensive description in \citet{Gudiksen_2011}. \\

We study a $12 \times 12 \times 12$ Mm$^3$ domain, where the vertical direction extends from a shallow, self-consistent convection zone ($\sim -2.5$~Mm) to the corona ($\sim 8$~Mm). The domain has a higher resolution than most published \verb|Bifrost| simulations, with 1024$^3$ grid points, leading to a horizontal resolution of $\mathrm{d}x=\mathrm{d}y=11.7$~km, and variable vertical resolution, $\mathrm{d}z$, between 6 and 35~km. We focus on a 36 minute window of time, starting 1.5 hours into the simulation (referred to as $t=0$~seconds from now on), with a snapshot cadence of 10 seconds. Motions in the convection self-consistently drive the dynamics further up in the box, and the boundary conditions are set such that no waves are artificially injected. To minimise interference from reflections at the boundaries, the horizontal boundary conditions are periodic, whilst the upper boundary uses characteristic boundary conditions \citep[see][]{Gudiksen_2011}. In \citetalias{Cherry_2025}, we took 4 horizontal slices through the domain $2$~Mm apart representing the photosphere (0~Mm), upper-chromosphere (2~Mm) and corona (4~Mm, 6~Mm). Although the full data set was analysed, the results of only a single snapshot in time were discussed. We now extend this analysis to 3D by considering a continuous $z$ direction, and by considering the temporal evolution of the simulation. 

\begin{figure}
    \centering
    \includegraphics[width=\linewidth]{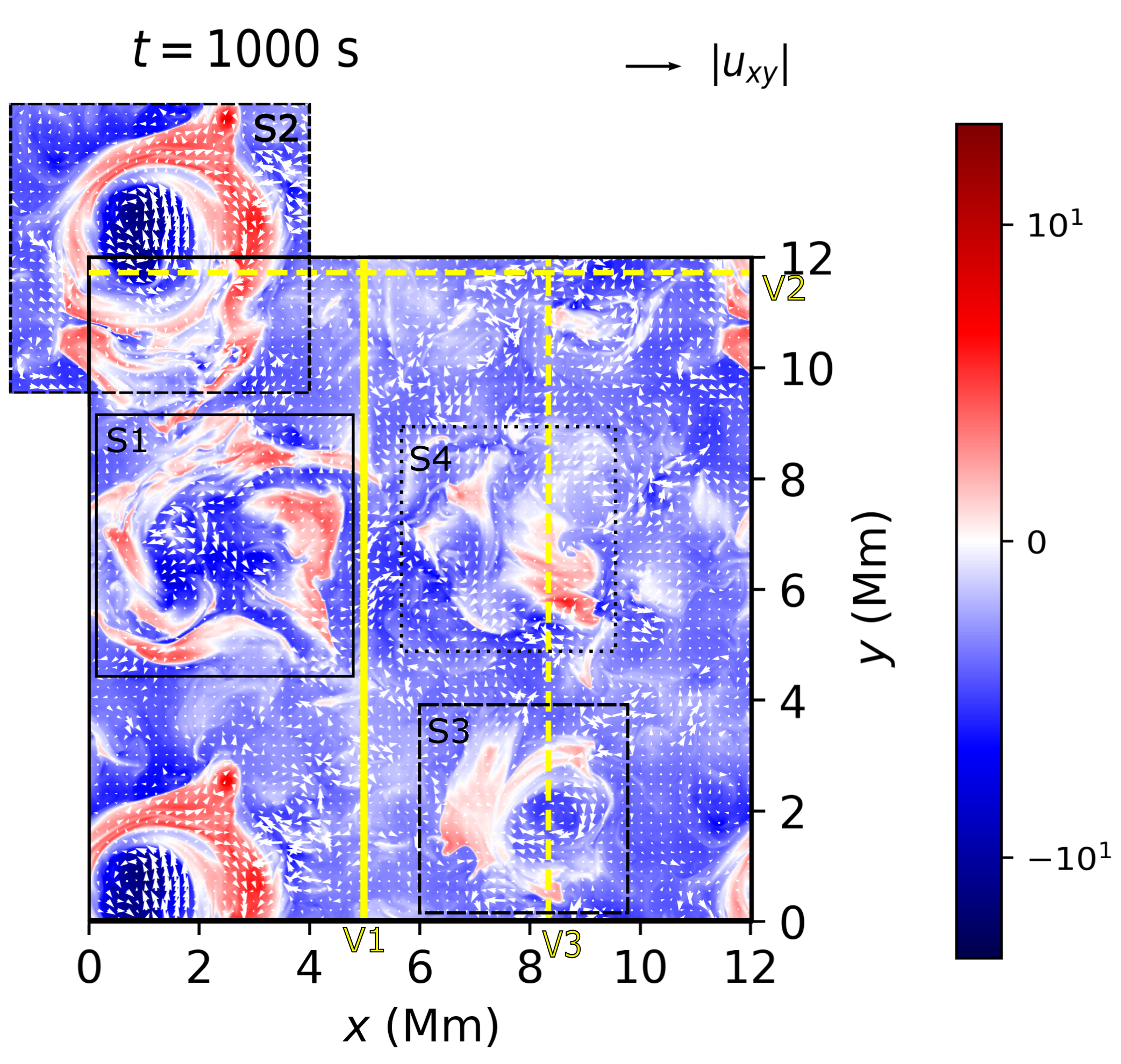}
    \caption{Vertical magnetic field at 2~Mm, with direction of horizontal velocity vector, $\mathbf{u}_{xy}$. We utilise the periodic boundaries to display the swirl S2, but note that the domain analysed in this study stays within the axis boundaries, as outlined by the solid black box. The vertical cuts through the domain used for analysis are displayed for reference by the solid and dotted lines, whilst the three main swirls are highlighted by the black boxes.}
    \label{fig: Swirl_slices}
\end{figure}

\subsection{Simulation characteristics}

\citet{Cherry_2025} discussed two prominent swirling features in the simulation, labelled as S2 and S3 in Figure \ref{fig: Swirl_slices}. A description of the qualities of these swirls is given in Table \ref{tab: swirls}, alongside two other swirls, S1 and S4. The swirl S1 will be the main focus of this study, since it is a large, long-lasting swirl that launches a single eruption, whilst S2 repeatedly erupts during the 36 minute window (see \cite{Finley_2022} for a detailed study on swirl evolution). The fourth swirl, S4, is outlined by the dashed yellow box in Figure \ref{fig: Swirl_slices}. This swirl is smaller and short-lived but interacts strongly with both S1 and S3.

%\begin{itemize}
    %\item \textcolor{red}{structure of large swirl: strong twisted vertical magnetic field surrounded by downwards magnetic field - See Quentin's plots.}
%\end{itemize}

\begin{figure}
    \centering
    \includegraphics[width=\linewidth]{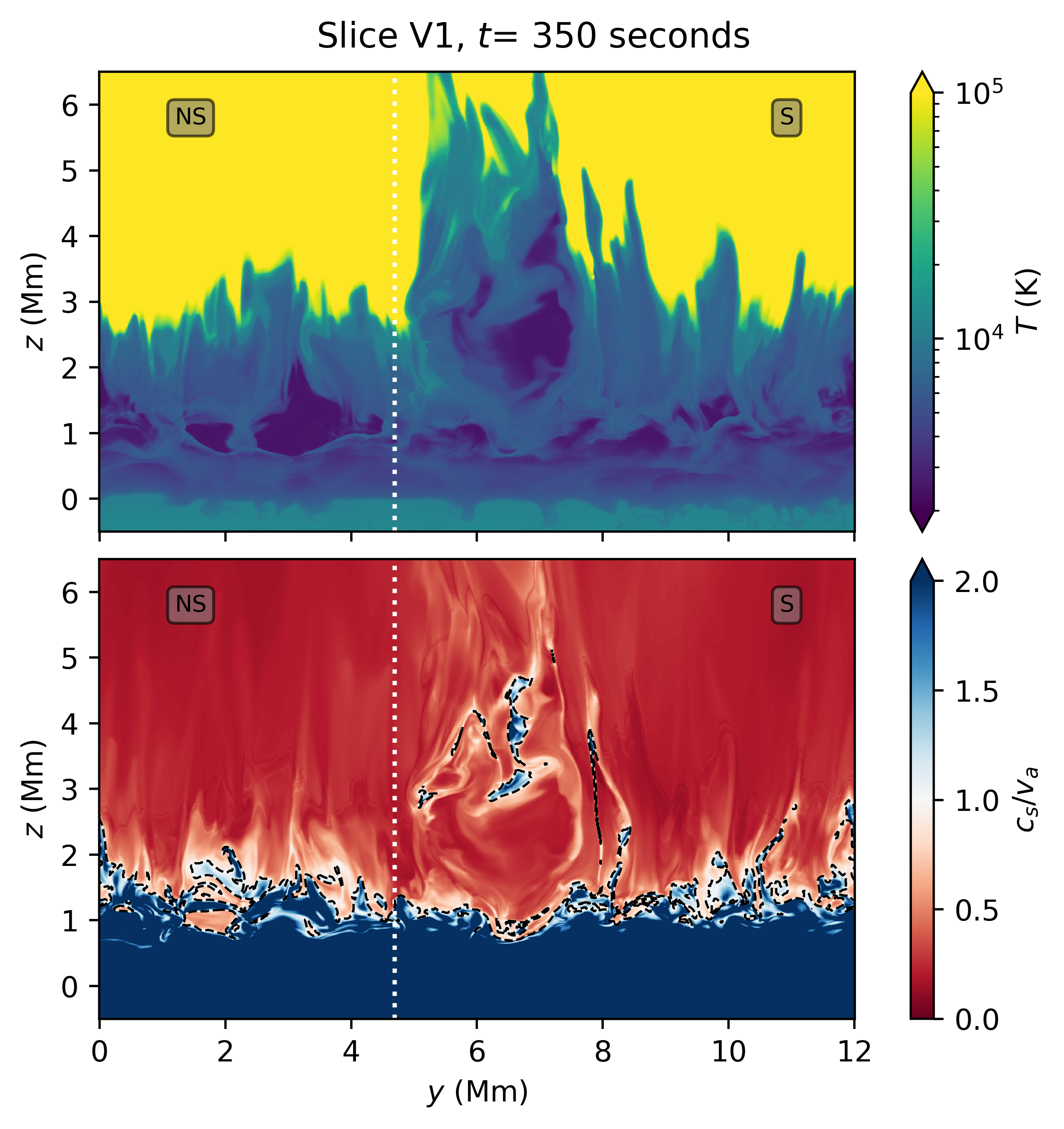}
    \caption{Upper: Temperature plot in vertical slice V1 at time $t=350$~s, where the slice intersects the swirl S1. We note the difference in time to Figure \ref{fig: Swirl_slices}. The scale is set to highlight the chromospheric plasma showing characteristics of type II spicules.
    Lower: Ratio of sound and Alfvén speeds in slice V1, with the equipartition level ($c_s/v_a = 1$) outlined by the dashed line. The vertical dotted line splits the slice into two sections (with swirl (S), without swirl (NS)), which will be used for analysis throughout Section \ref{sec: results}.}
    \label{fig: spic}
\end{figure}

In the absence of swirls, we see a complex field consisting of jets characteristic of type I spicules (Figure \ref{fig: spic}, upper: see \cite{Sykora_2017} for a detailed study on spicules in numerical simulations).

\begin{table}
    \centering
    \begin{tabular}{c|c|c|c|c}
        Label & area (Mm$^2$)& start t (s) & duration (s) & height (Mm)\\
        \hline \hline
        S1  & 20 & 240 &  $>1500$ & 6 \\
        S2* & 25 & 260 & $>1500$&  8 \\
        S3  & 4 & 550 & $>1000$&  6 \\
        S4* & 4  & 750 & 400 &  3 \\
    \end{tabular}
    \caption{Characteristics of selected swirls in the domain in Figure \ref{fig: Swirl_slices} used for analysis. Starred swirls have multiple eruptions. The area is taken at $z=2$~Mm, whilst height refers to the height from the photosphere to the maximum height of the swirling structure.}
    \label{tab: swirls}
\end{table}

\subsection{Vertical slices}
Figure \ref{fig: Swirl_slices} shows three vertical slices in the domain, labelled V1-3.These slices are chosen since they intersect the largest-scale swirls in the domain, as discussed above. Even in this quiet simulation, it is difficult to find regions of plasma which are truly quiet - exhibiting no interactions with swirling structures. For the results of this study, we focus on slice V1, which is unique in that it can be split into two regimes exhibiting characteristics of swirling (S) and non-swirling (NS, quiet) plasma (see Figure \ref{fig: spic}). Swirl S1 intersects V1 from time $t=230$-600~s, and borders the swirling plasma of both S1 and S4 for the remainder, capturing the interactions between the swirls. The non-swirling regime is taken for $y<4$~Mm. since the there is about 1~Mm either side of the slice that does not contain plasma which interacts with the swirls S1 annd S3. We have also considered V2 and V3 in our analysis, as slices which intersect the other large swirls discussed above. These slices focus more on the inner mechanisms of the swirl structures, and have little to no quiet plasma. We discuss these slices in comparison to V1 in Section \ref{sec: swirl_discuss}. See the supplementary files for a video of the temporal evolution of Figures \ref{fig: Swirl_slices} and \ref{fig: spic}.

%%%%%%%%%%%%%%%%%METHODS%%%%%%%%%%%%%%%%%%%%%%%%%%%%%%%%%%%%%%%%

% \begin{figure}[h]
%     \centering
%     \includegraphics[width=\linewidth]{Images/beta.png}
%     \caption{Ratio of sound and Alfvén speeds in slice V1, with the equipartition level ($c_s/v_a = 1$) outlined by the dashed line. The vertical dotted line splits the slice into two sections (with swirl, without swirl), which will be used for analysis throughout Section \ref{sec: results}.}
%     \label{fig: beta}
% \end{figure}

\section{Methodology}
\label{sec: Methedology}
To detect wave activity in localised regions of the domain, we use the Most Significant Frequency (MSF) method, as described in \citetalias{Cherry_2025}. This method is an extension to the Discrete Fourier Transform (DFT),
\begin{multline}
     \mathcal{F}(\tilde{\nu}_x,\tilde{\nu}_y,f) = \\
    \sum_{n_x = 1}^{N_x} \left(\sum_{n_y = 1}^{N_y} \left(\sum_{n_t = 1}^{N_t} g(n_x,n_y,n_t)e^{-i \tfrac{ 2\pi f n_t}{N_t}} \right) e^{-i \tfrac{ 2\pi \tilde{\nu}_x n_x}{N_x}} \right) e^{-i\tfrac{2\pi \tilde{\nu}_y n_y}{N_y}},    
\end{multline}
with frequency ($\omega$) and wavenumbers ($\tilde{\nu}_x$,$\tilde{\nu}_y$), where $N_x$, $N_y$, and $N_t$ are the number of horizontal grid points in $x$ and $y$, and the total number of snapshots, respectively. This gives us a maximum frequency of 50~mHz and a resolution of 0.526~mHz. \\

The MSF algorithm selects the individual frequencies with the closest signal output from the inverse DFT to the original data at each point. Through this method we can discover dominant frequencies in localised regions of the domain, as described in \citetalias{Cherry_2025}. To balance accuracy and computational speed, we choose to compute the MSFs for individual frequencies up to 19.47~mHz (the first 38 frequencies). The final bin includes the signal from all other frequencies (up to 50~mHz). Our choice in frequency range in the MSF decomposition allows the final frequency bin to capture sudden gradients such as swirl boundaries and shock fronts, too (see Section \ref{sec: shocks}). It must be noted that this is merely a useful consequence of the specific frequency range chosen and lower frequencies would, in most regions, replace the highest frequency (HF) MSFs in a MSF decomposition with a larger frequency range (see Appendix \ref{app: high_freq}). However, it was considered beneficial, both in terms of computational speed and the detection of shocks, not to extend the range in this case. The exception to this is when the MSF decomposition is applied to the velocity perpendicular to the magnetic field, $u_\perp$. In this case, the result sensitivity required a larger frequency range in order to capture any physical frequencies. Therefore, the highest individual frequency considered in the MSF calculation for $u_\perp$ (i.e., the upper limit of the frequency range) is 50~mHz, compared to 19.47~mHz for the other variables.\\

\section{Results} \label{sec: results}
\begin{figure}[h]
    \centering
    \includegraphics[width=\linewidth]{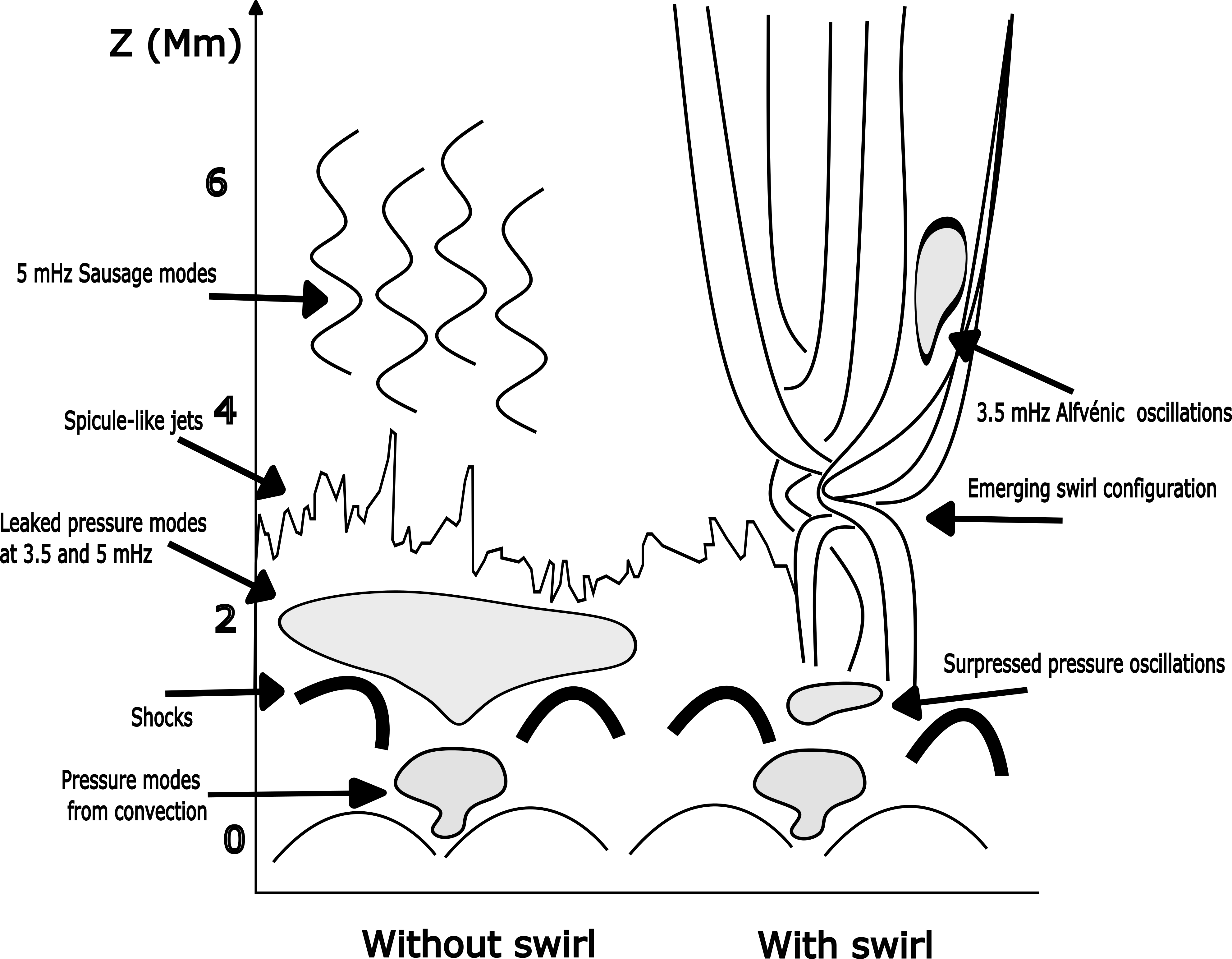}
    \caption{Sketch and general overview of results, including suggested mode classification.}
    \label{fig: sketch}
\end{figure}

The vertical slice V1 is analysed from the photosphere (0~Mm) to corona (6.5~Mm). We omit the top 1.5~Mm of the simulation box in order to minimise the effects of reflection due to imperfections in the top boundary condition. Figure \ref{fig: spic} outlines the structures in slice V1, through the temperature (top panel) and the ratio of the acoustic and Alfvén speeds (bottom panel). The slice contains quiet (spicule-like) conditions, as well as a twisted swirl structure, which pushes the equipartition level to coronal heights. Naturally, this allowed us to examine the MSFs versus height in swirling  plasma that interacts with twisted swirl structures and non-swirling plasma (case S and NS, respectively, shown in Figure \ref{fig: spic}). We found that the presence of a swirl drastically changes the behaviour of dominant wave propagation, and hence the energy that can be transported to the corona from the lower atmosphere.  This result is shown schematically in Fig. \ref{fig: sketch}.

%%Results

\subsection{Global oscillations} \label{sec: glob_osc}

\begin{figure}[h]
    \centering
    \includegraphics[width=0.9\linewidth]{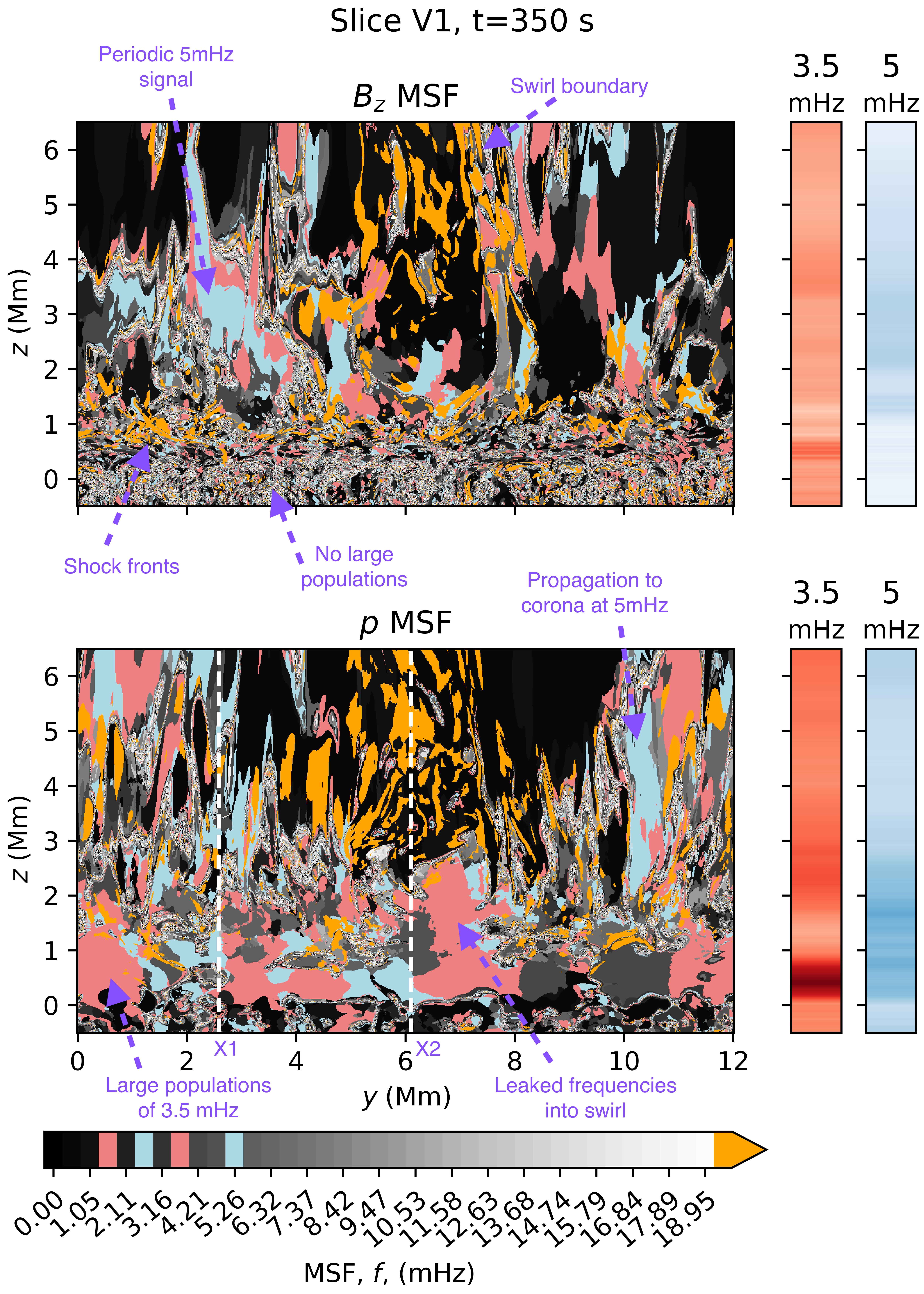}
    \caption{Vertical slice V1, identified in Figure \ref{fig: Swirl_slices}, at $t=350$ s. MSF results for the vertical magnetic field, $B_z$ (top), and pressure, $p$ (bottom), are shown. The dominant frequencies 3.6 and 5.2~mHz are highlighted. It is considered that these frequencies are also captured in their half-frequency counter parts, and so these are also highlighted. Finally, the HF bin, $f \geq 19.47$~mHz is highlighted as an example of capturing the steep gradients in the domain, such as shocks around $z=1$~Mm and the swirl structure. The bars to the side show the percentage of the domain covered by each frequency vs. height, with the darkest shades representing the most coverage ($>$35$\%$), and the lightest representing zero coverage. These quantities are expanded in Figure \ref{fig: Coverage_swirl}. The two slices used in Figure \ref{fig: t-d_upar_p} are shown by the dashed lines.}
    \label{fig: MSF_vert}
\end{figure}

\begin{figure*}[h!]
    \centering
    \includegraphics[width=0.9\linewidth]{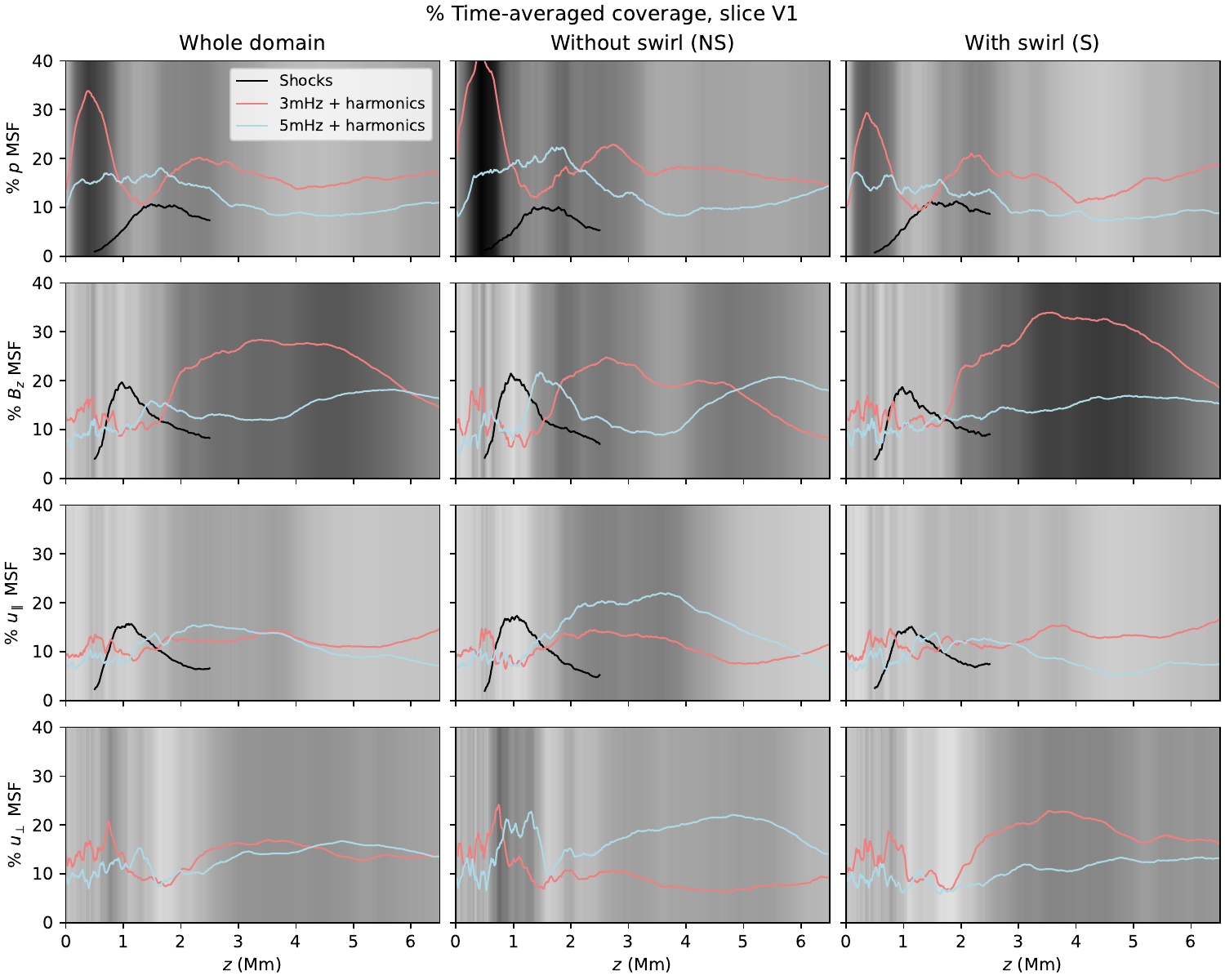}
    \caption{Time-averaged coverage of vertical slice V1 of $f= 3.5$ and 5 mHz oscillations, including harmonics at 1.5~mHz, and 2.5~mHz respectively, with height. The HF bin is included in black as a proxy to the coverage for shocks in the region 0.5-2.5~Mm. It is noted that the HF bin for $u_\perp$ is not included, since only individual frequencies are considered. Underneath, the total domain coverage by the two MSFs (3.5 and 5~mHz) is shown, with the darkest regions representing the most coverage ($>60\%$) and the white regions representing no coverage.}
    \label{fig: Coverage_swirl}
\end{figure*}

\begin{figure*}[h!]
    \centering
    \includegraphics[width=0.9\linewidth]{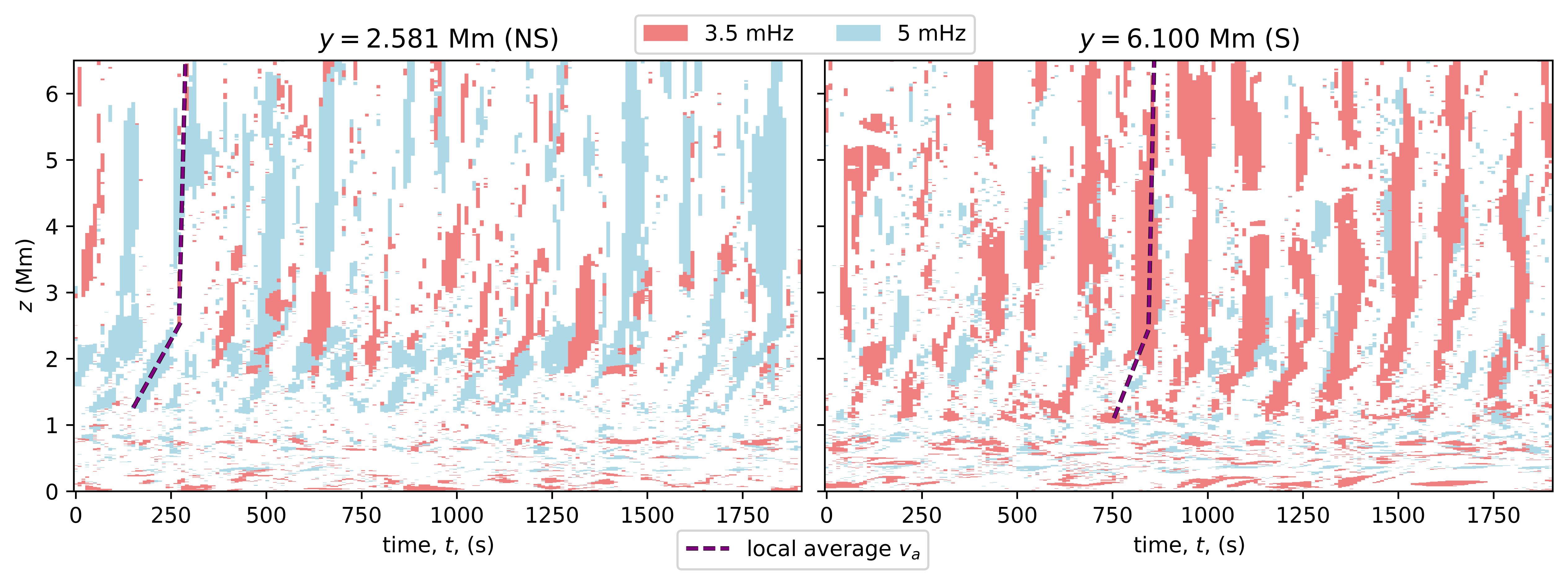}
    \caption{Time-distance plot of vertical slice V1 at $y=2.6$~Mm (left panel, case NS) and $y=6.1$~Mm (right panel, case S). See figure \ref{fig: MSF_vert} for slice references. MSFs of 3.5 and 5 mHz are displayed for parallel velocity, including harmonics. The dashed lines over-plotted represent the trajectory of an object moving at the Alfvén speed. For this, the average local Alfvén speed is calculated for the specified times for $z=1.5$-2.5~Mm, and for  $z=2.5$-6.5~Mm.}
    \label{fig: t-d_upar_p}
\end{figure*}

In \citetalias{Cherry_2025}, we identified the global dominant frequencies for multiple horizontal slices in the photosphere to corona to be in the range 3-6~mHz. This result was found using the power spectrum density and global coverage of the MSFs (in the form of pixel counts) for the four horizontal planes at a single time, $t$. In this study, we focus on the evolution and vertical propagation of the global oscillations at 3.5 and 5 mHz, whose multiplicities are highlighted in Figure \ref{fig: MSF_vert}. Here, we display MSFs applied to the vertical magnetic field, $B_z$, and pressure, $p$ at time $t=350$~s, where the vertical slice intersects the swirl. We have chosen these variables to represent magnetic (transverse) and acoustic (longitudinal) fluctuations, respectively, as described in \citetalias{Cherry_2025}. Alongside these variables, we shall also consider the velocity parallel to the magnetic field, $u_\parallel$, for longitudinal oscillations, and the velocity perpendicular to the magnetic field, $u_\perp$, for magnetic  fluctuations. It is, in fact, MSFs at 3.6~mHz and 5.3~mHz which dominate both variables. However, due to the frequency step size, the corresponding bins encompass 3.5 and 5 mHz respectively, and as such for simplicity we refer to these MSFs as 3.5 and 5 mHz frequencies from now on. We found that the MSF algorithm also detected these frequencies in the 1.6 and 2.6~mHz range, so we have considered these as harmonics of the 3.5~mHz and 5 mHz frequencies, respectively, and included them in our analysis. These two frequencies cover a significant portion of the domain at most heights. In the pressure (Figure \ref{fig: MSF_vert}, bottom panel), it is clear to see the generation of large signals at 3.5 and, to a lesser extent, 5~mHz frequencies in the photosphere. The locations of these 3.5~mHz frequencies in the photosphere align with the granules, showing them to be the driver of these oscillations. Only some of the frequencies leak upwards into the chromosphere and onwards to the corona. There is a distinct lack of these frequencies in the upper atmosphere surrounding the swirl (outlined by the HF bin in orange) in the upper atmosphere, although some 3.5~mHz frequencies have leaked into the foot point of the swirl near $z=2$~Mm. On the contrary, in the regions further away from the swirling structure ($y<4$~Mm and $y>10$~Mm), both 3.5 and 5~mHz oscillations can be found extending up to and above $z=6$~Mm. The vertical magnetic field (top panel), shows both dominant frequencies found in large, continuous clusters starting between $z=1$-2~Mm. Between $y=2$-4~Mm, there is a strong signal at 5~mHz. In time, this signal periodically appears around $z=2$~Mm, rises slightly upwards to $4$~Mm, whilst being advected towards the swirl, before the majority sinks back down to 2~Mm, whilst some oscillations are pushed further up to 6~Mm. In contrast, there is a larger coverage of 3.5 mHz on the boundaries of the swirl. It seems that both variables show notable differences in the S and NS regimes. \\

To investigate this further, we define the coverage, $C$, of each frequency at height $z$ in the vertical plane V1 as the percentage of the $N_x$ grid cells covered by each MSF, 
\begin{equation}
    C_f(z) = N_f(z)/N_x * 100,
\end{equation}
where $N_f$ is the number of points with MSF $f$ at height $z$. The time-averaged coverage for 3.5 and 5~mHz (from $t=100-1800$~s to avoid boundary discrepancies in the temporal DFT) for the total domain are displayed in the rightmost plots of Figure \ref{fig: MSF_vert} and split into the two cases, S $\&$ NS, in Figure \ref{fig: Coverage_swirl} for all variables. Immediately, it is clear to see the difference in the two cases. In the variables associated with transverse magnetic fluctuations, $B_z$ and $u_\perp$, the 5~mHz frequencies dominate over the 3~mHz in the upper atmosphere in case NS. Nevertheless, these frequencies are suppressed in case S, allowing the 3~mHz frequencies to dominate. As for the acoustic fluctuations, we see a reservoir of 3~mHz MSFs in the photosphere for pressure, covering up to 40$\%$ of the domain. Although present in both cases, this reservoir decreases by up to 10$\%$ in case S. \\

From around $500$~km -- the temperature minimum and the start of the chromosphere -- there is a steep decline in the 3.5~mHz signal in pressure, leading to a dip (local minimum) in the coverage between $z=1$ and 2~Mm. The decrease in coverage is correlated with the rapid formation of shocks in the chromosphere, represented here by the HF bin, and discussed more in Section \ref{sec: shocks}. From $2$~Mm, there is a small resurgence in the pressure signal at 3.5~mHz in both cases. Above this, and only in case S, the parallel velocity exhibits a response to the resurgence through a large enhancement of 3.5~mHz MSFs above 2~Mm. There is also a response in the perpendicular velocity, characterising a transverse component to the mode, enabled by the swirl configuration. We assume this is due to the untwisting, circular motions of the swirls, as they have components of velocity parallel and perpendicular to the magnetic field. Through the perpendicular and parallel velocities, the 3.5~mHz frequencies reach the corona and fill a significant area of the S region (25-35$\%$). Similar reservoir-response behaviour is observed in the 5~mHz frequencies in case NS, although, crucially, without the same dip at $z=1$-2~Mm as seen in the 3.5~mHz signal. This behaviour suggests a significant connection between acoustic signals in the lower atmosphere and magnetic signals in the upper atmosphere, via a mode conversion mechanism around 2~Mm.\\

Figure \ref{fig: t-d_upar_p} clearly captures the vertical propagation of the parallel velocity MSFs in time along vertical slices $y_1=2.6$~Mm and $y_2=6.1$~Mm (as displayed in Figure \ref{fig: MSF_vert}), relating to case NS and case S, respectively. In case S, propagation is dominated by 3.5~mHz signals from 1.5~Mm at the Alfvén speed (marked by the dashed lines), whilst in case NS, the same is true with 5~mHz signals. At $2.5$~Mm, there is a distinct increase in the propagation speed, which directly relates to the level at which the Alfvén speed also increases significantly. Therefore, these modes are primarily magnetic from at least $1$~Mm upwards. A similar behaviour is shown by the 5~mHz pressure oscillations in case NS, generally, with continuous propagation from the chromosphere ($\sim$ 500~km) to the corona (5~Mm). This suggests a magnetoacoustic mode, such as a sausage or kink mode, which can be determined through further analysis and is discussed in Section \ref{sec: mode}. In case S, however, the propagation of 3.5~mHz pressure oscillations stops abruptly just above $2$~Mm. This corresponds with the emergence of the same frequency in the vertical magnetic field, suggesting a transition from acoustic to magnetic oscillations at this height.

\subsection{Energy transfer} \label{sec: energy}

\begin{figure}[h!]
    \centering
    \includegraphics[width=\linewidth]{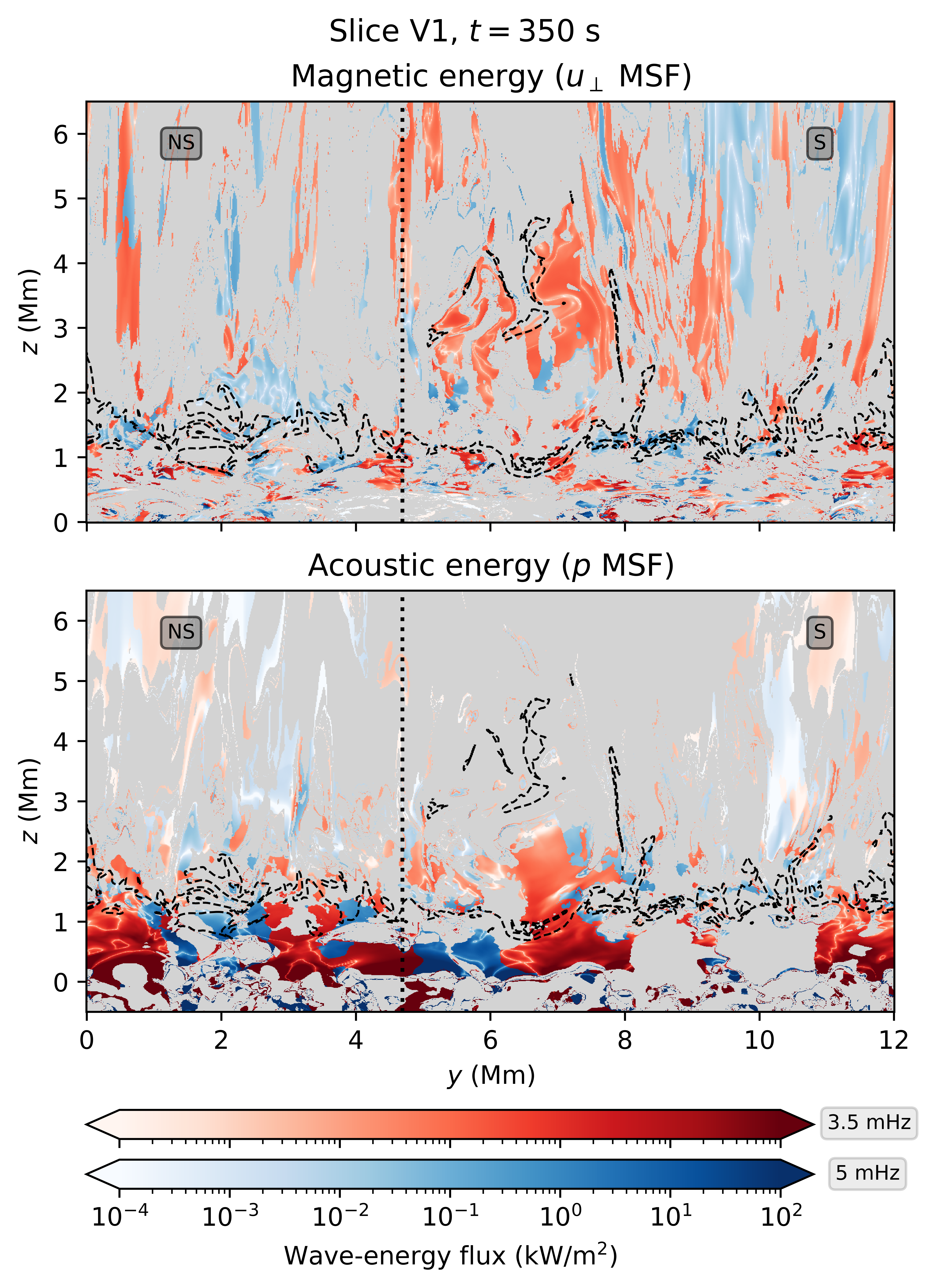}
    \caption{Instantaneous wave-energy flux in slice V1 at $t=350$~s. The absolute magnitude of the magnetic flux (upper panel) is calculated in the locations of 3.5 and 5~mHz signals in perpendicular velocity, whilst the acoustic flux (lower panel) is calculated in the locations of 3.5 and 5~mHz signals in pressure. The equipartition level is overlaid as a dashed line. The dotted vertical line represents the partition between case S and case NS.}
    \label{fig: energy_loc}
\end{figure}

We calculate the vertical wave power through height $z$ for acoustic and magnetic waves in the regions with MSFs at 3~mHz and 5~mHz, as demonstrated in Figure \ref{fig: energy_loc}, and compare to the power in all frequencies (i.e., in the full domain). We calculate the acoustic wave-energy flux defined for longitudinal acoustic waves travelling vertically through the domain,
\begin{equation}
    F_\text{a} = \frac{1}{2}\rho v_z^2 c_s, \label{eq: ac}
\end{equation}
at regions with the specified MSFs in pressure and the magnetic wave-energy flux defined for vertically-propagating Alfvénic waves driven by horizontal shaking motions by the vertical shearing Poynting flux,
\begin{equation}
    S_{z,s} = \frac{-B_z}{\mu_0}(v_xB_x + v_yB_y), \label{eq: mag}
\end{equation}
at regions with the specified MSFs in velocity perpendicular to the magnetic field. \\

In the acoustic flux (Equation \ref{eq: ac}), we note that one could use $v_\parallel$ instead of $v_z$ for a better approximation to the total flux travelling upwards along the field lines, but in our case the difference was negligible and we are most interested in vertical wave propagation. We also note that for vertical wave-energy flux associated to fast magnetoacoustic modes, one may use the vertical emerging Poynting flux,
\begin{equation}
    S_{z,e} = \frac{v_z}{\mu_0}(B_x^2 + B_y^2).
\end{equation}
However, this term was found to be much smaller than the shearing Poynting flux in this study, since fast magnetoacoustic modes primarily travel perpendicular to the field lines, hence horizontally in the domain. We choose the regions of MSFs in p and perpendicular velocity as the most consistent markers of acoustic and magnetic waves, respectively. However, we note that regions in parallel velocity and the vertical magnetic field could also be used. \\

The acoustic wave power at height $z$ is then defined as the total acoustic flux in the specified area:
\begin{equation}
    P_a = \iint F_a (x,y,t) \, \mathrm{d}x\mathrm{d}y,
\end{equation}
and similarly for the magnetic power. \\

\begin{figure*}[h!]
    \centering
    \includegraphics[width=\linewidth]{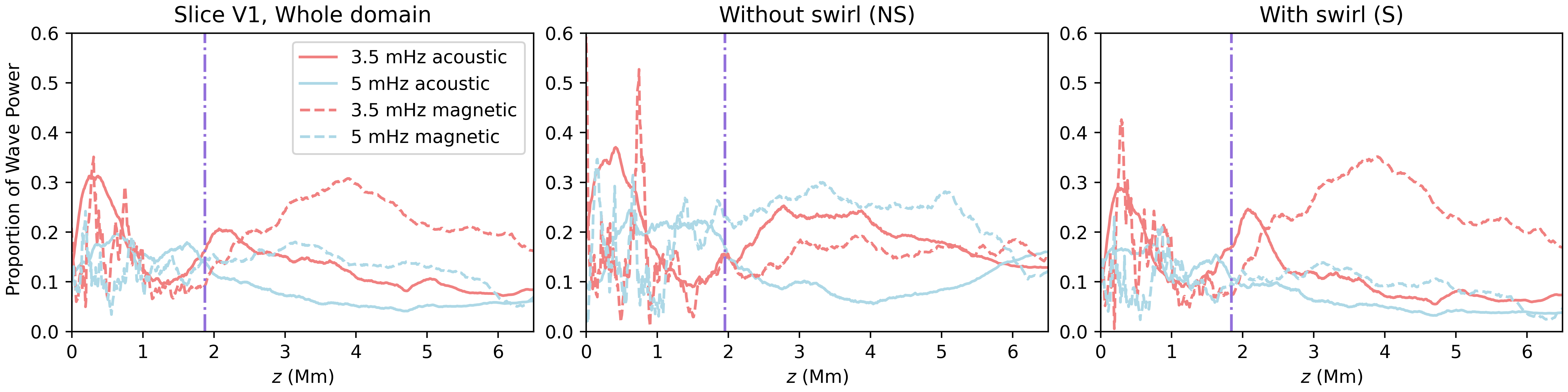}
    \caption{Proportion of the total time-averaged vertical wave-energy power found at locations of 3.5 mHz and 5 mHz MSFs in the whole domain V1 (left panel) and in the split domain (NS, S) for the presence and absence of a swirl structure (middle and right panels). The dash-dot lines represent the height at which the total magnetic power  surpasses the acoustic power in the domain.}
    \label{fig: wave-energy_avg}
\end{figure*}

The total time-averaged wave power across all frequencies amounts to $10^{14}$~kW acoustic power in the photosphere, exponentially dropping to  $10^8$~kW in the corona, and $10^{11}$~kW magnetic power in the photosphere, dropping to $10^9$~kW in the corona. In each case, the magnetic wave power surpasses the acoustic power at $z\approx 2$~Mm (see Figure \ref{fig: wave-energy_avg}), with this transition occurring lower in case S than in case NS. In this figure, we also display the proportion of the total wave power described above which is generated in regions with specifically 3.5 and 5~mHz MSF signals. The average contribution to the overall power in the domain lies between 10 and 20$\%$ for both frequencies, with a combined average of $30\%$, composed of two-thirds dominant power to one-third non-dominant either side of the transition. It is clear to see the influence of the coverage of each frequency (Figure \ref{fig: Coverage_swirl}), especially in the reservoir-resurgence behaviour of the acoustic power at 3.5~mHz. Here, the power contribution is directly correlated to the coverage of the domain. There are peaks in the individual magnetic power reaching over 30$\%$ for both 3.5~mHz signals in case S, and 5~mHz  signals in case NS. At these peaks. the combined power constitutes over 50$\%$ of the power in all frequencies. The coverage, however, stays around $20\%$ for both signals at these heights, respectively, showing an enhancement of power compared to regions of other MSFs. The enhancement in the magnetic power can also be seen in the full domain (left panel of Figure \ref{fig: wave-energy_avg}) and is a response to the resurgence in acoustic power around $z=2$~Mm, despite the full domain coverage of perpendicular velocity not detecting such a strong response. This is coupled with the large response seen in the coverage of MSFs parallel velocity. \\

This magnetic response at 3.5~mHz is not seen in the quiet plasma of case NS (middle panel). Here, the acoustic power contribution at 3.5~mHz remains larger for longer above $z=2$~Mm, instead of decaying as in the full domain and case S. There is no considerable difference in the pressure coverage at 3.5~mHz between the different cases, so once again this suggests an enhancement of power. This enhancement of acoustic power at 3.5~mHz coincides with the enhancement of magnetic power at 5~mHz, whilst the 3.5~mHz magnetic power, which is enhanced in the full domain and case S, is suppressed with a $10\%$ difference in contribution to the full domain. Therefore, the magnetic power at 5~mHz is the dominant power from $z=2-6$~Mm, making up an average of 23$\%$ of the total vertically-propagating magnetic power. \\

\begin{figure}
    \centering
    \includegraphics[width=\linewidth]{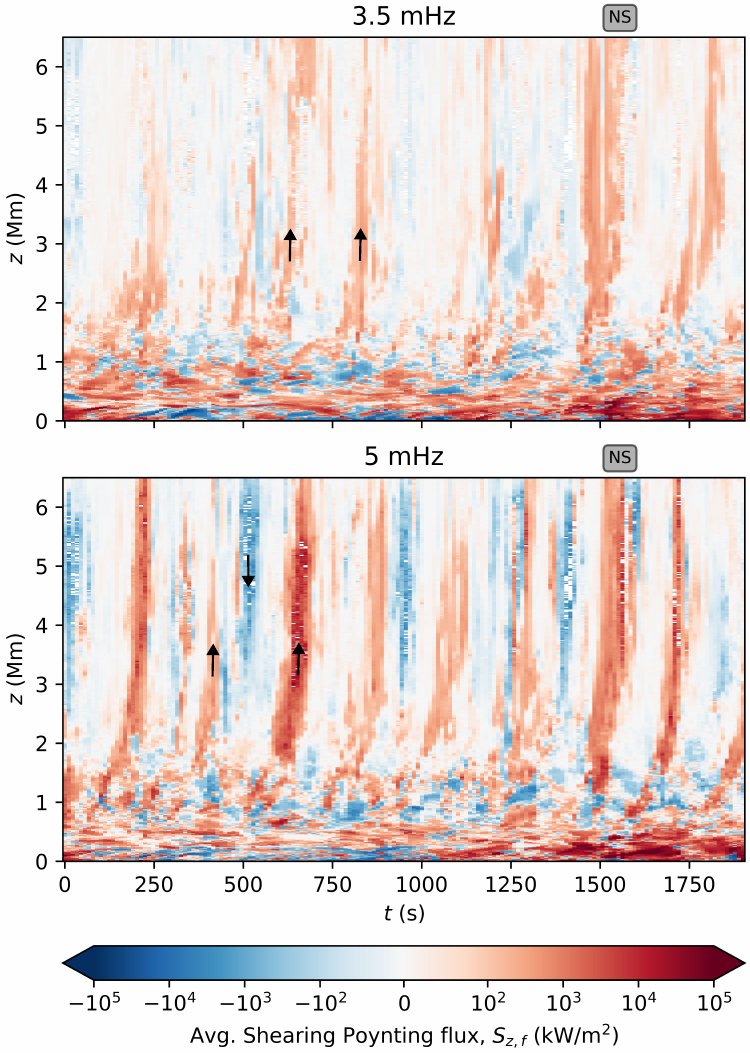}
    \caption{Horizontally-averaged shearing Poynting flux in the NS region of slice V1 for 3.5 (top) and 5 (bottom)~mHz. The arrows show the direction of energy propagation.}
    \label{fig: Poynt_flux}
\end{figure}

The 3.5~mHz frequencies in case NS produce a further $15\%$ of the total magnetic power, whilst covering half as much of the domain as the 5~mHz MSFs. We note here that the magnetic power is calculated from the net magnetic wave-flux of both downward and upward propagating signals. Therefore, we can infer from the power comparison of 3.5~mHz and 5~mHz signals that the former are more efficient at reaching and depositing energy in the corona, whilst the latter also contain strong downwards propagation. By taking a horizontal average of the Poynting flux across the region of case NS, as in Figure \ref{fig: Poynt_flux}, it is indeed clear to see relatively much stronger down-flows at 5~mHz than 3.5~mHz in time. Interestingly, downwards propagation in the 3.5~mHz signals is found lower down in the atmosphere (below $z=3$~Mm), where as down-flows of 5~mHz power are strongest between $z=3-6$~Mm. Therefore, the power that reaches the corona from 3.5~mHz signals is also deposited there, whilst the circulation of power from 5~mHz signals  limits the deposition into the upper atmosphere.

\subsection{Shocks} \label{sec: shocks}

In both the coverage and the contribution to the wave power of 3.5~mHz MSFs, there is a characteristic dip in the values between $z=1-2$~Mm -- the chromosphere. In this region there is a high prevalence of shock activity, as waves experience the rapidly decreasing sound speed and density. The propagation of magneto-acoustic shocks is characterized by high spatial and temporal frequencies due to the presence of sharp gradients. Consequently, in the application of the MSFs, shock-dominated areas should be captured by regions exhibiting high-frequency components in the temporal domain. In this sense, it is logical to assume that the HF bin in our results will contain such activity, as we did in Figure \ref{fig: Coverage_swirl}. However, the turbulent and multi-scale dynamics present in our model also give rise to linear wave modes that span a wide range of frequencies. We must therefore keep in mind that the high-frequency regime resulting from the MSF decomposition may contain both shocks and high-frequency linear waves. \\

In this context, our aim in this section is to establish that the signals identified in Figure~\ref{fig: Coverage_swirl} can be interpreted as a manifestation of magneto-acoustic shock dynamics. To this end, we adopt the methodology developed in Noraz et al. (2025, submitted) (see also \cite{Yokoyama_2020} and \cite{Finley_2022}) to perform a shock identification using the sonic-compression criterion. A grid point is classified as being within a shock if:
\begin{equation}
        -\nabla\cdot\mathbf{v} > \frac{c_s}{\epsilon \, \mathrm{d}s},
        \label{eq:cs_crit}
\end{equation}
where $c_s$ is the local sound speed derived from the tabulated EOS, and $\mathrm{d}s=\max(\mathrm{d}x,\mathrm{d}y,\mathrm{d}z)$. We adopt $\epsilon=6$, which yields a highly selective threshold to avoid contamination from high-amplitude linear waves. This conservative criterion hence likely excludes broad or low-amplitude shocks, but ensures that identified regions represent unambiguous shock features (see discussion in Noraz et al. 2025, submitted).\\

In a second step, we classify the detected shocks as either fast or slow. We know that fast shocks increase the magnetic field strength, whereas slow shocks lead to its reduction (see \cite{Priest2014} and references therein). Since all collisional shocks necessarily induce local compression and an associated pressure enhancement, we compute the local divergence of the pressure, $\nabla (p)$, and that of the magnetic energy density, $\nabla (B^2)$, at each identified shock location. The sign of their product allows us to determine shock types: fast if positive, slow if negative.\\

\begin{figure}
    \centering
    \includegraphics[width=\linewidth]{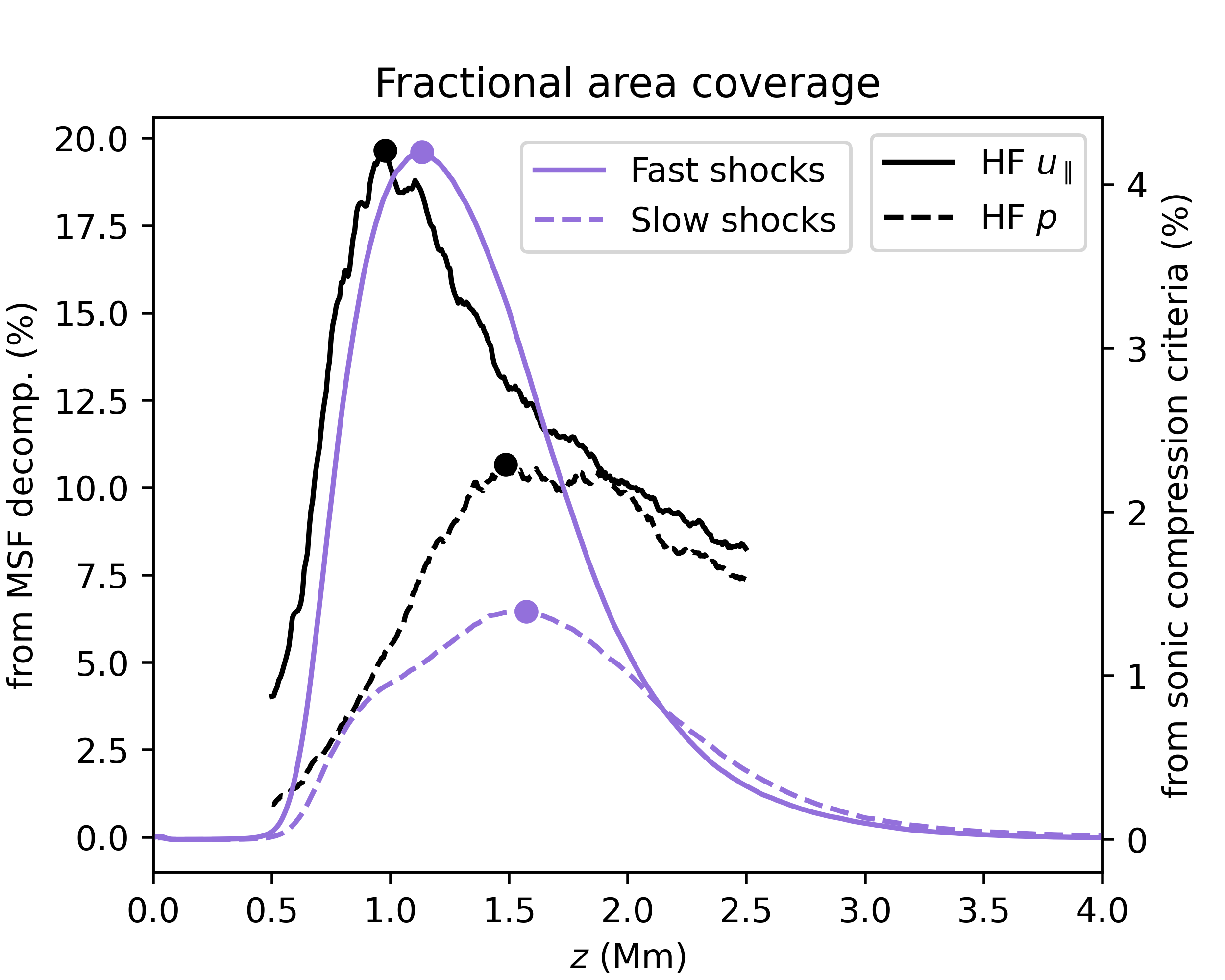}
    \caption{Fast \textit{vs.} Slow shocks distinction. Fractional area coverage over the whole domain as a function of height, as deduced from the sonic-compression criteria (Equation~\ref{eq:cs_crit}, magenta) and the HF bin ($\geq 19.47$~mHz) of the MSF decomposition (black). Note the different ordinate axis on both sides of the Figure.}
    \label{fig:FastSlow}
\end{figure}

Figure~\ref{fig:FastSlow} then presents the fractional area covered by fast (solid violet) and slow (dashed violet) shocks as a function of altitude, deduced from the sonic-compression criteria Eq.~\ref{eq:cs_crit}. We compare these with the fraction of area selected via the HF bin of the MSF decomposition, applied to the parallel velocity, $u_\parallel$, and pressure, $p$, fields (see Section~\ref{sec: glob_osc}). We note that (1) the MSF signal in $u_\parallel$, peaking near z=0.98~Mm, aligns closely with the peak of fast shock coverage at $z=1.13$~Mm, but also that (2) the MSF signal in pressure oscillations peaks at $z=1.48$~Mm, which also coincides closely with the peak of slow shock detection at $z=1.57$. While the amplitude of fractional coverage differs between the MSF-based (black) and shock-criterion-based (violet) curves, which reflects the inclusion of linear waves in the former and conservative character of the latter, the qualitative agreement between their spatial profiles supports the physical interpretation that the MSF-$u_\parallel$ signal primarily traces fast magneto-acoustic shocks, which convert energy from the kinetic reservoir into the magnetic one, while the MSF-$p$ signal is predominantly associated with slow shocks. \\

We note that this is not a quantitive conclusion and relies on the HF threshold -- that is, the largest frequency treated individually by the MSF method, which in this case is $19$~mHz. The results of the perpendicular velocity, for example, which calculate all frequencies individually (so has a threshold of 50~mHz), no longer resembles the slow shock curve, instead having only a narrow peak at $z=1.78$~Mm. In the same way that changing $\epsilon$ tunes the sonic-compression criterion, the HF threshold tunes the MSF for shock detection, and therefore this behaviour needs to be quantitively assessed for confident results. Never-the-less this is a poignant feature of the MSF decomposition. For our results in discussed in Section \ref{sec: glob_osc}, it is important to acknowledge that whilst there is a strong correlation between the peak of the highest frequencies and the 3.5~mHz dip, shocks cannot account for the entire decrease, since the signal captured by the sonic-compression algorithm is around 5 times smaller (for fast shocks). Even with a less strict $\epsilon$, it is most likely that another factor causes the dissipation of a significant fraction of the 3.5~mHz frequencies in this chromosphere, such as the acoustic cut-off frequency.

%\subsection{\textcolor{red}{Damping mechanisms}}

%%Discussion
\section{Discussion and conclusion}\label{sec:concl}
Using the MSF decomposition, we have been able to track and analyse the upwards propagation of two dominant frequencies, 3.5 and 5~mHz, in a realistic simulation of the solar atmosphere. Using a vertical slice, we split the domain into two regions. The first region (NS) contains quiet plasma displaying behaviour typical of type I spicules, whilst the other region (S) contains a swirl structure, which emerges near the beginning of the time series. We find that the MSF method captures some of the dominant frequencies as their half-equivalents, $1.6$ and $2.5$~mHz, and so include these location in the analysis as well. We note that the harmonics at $7$~mHz and 10~mHz also seem to coincide with the locations of these dominant frequencies, but here we do not include them, as their abundance is negligible.

\subsection{Mode determination} \label{sec: mode}

There is a reservoir of pressure oscillations at 3.5~mHz at photospheric levels, suggesting these correspond to pressure modes, driven by the granulation. At around $z=1$~Mm, the large decrease in the 3.5~mHz signal corresponds to an increase in fast magnetoacoustic shocks at this height (see Figure~\ref{fig:FastSlow}). There is a small resurgence in pressure oscillations of this frequency towards $z=2$~Mm. We speculate that the 3.5~mHz pressure modes in the photosphere form shocks due to the exponentially decreasing density at $z=1$~Mm, before reaching the high chromosphere and passing through the transition region. Due to the sudden increase in sound speed, the shocks can decompress to form pressure modes again as they reach the transition region. The resurgence abundance is smaller than the original reservoir due to the strong dissipation of energy in shock fronts, as well as the reflection of pressure modes due to the decreasing density and evanescence due to the acoustic cut-off frequency. Above 2~Mm, we see a large response of 3.5~mHz oscillations in the velocity parallel to the magnetic field, alongside a steep increase in the magnetic wave power at this frequency. This suggests a mode conversion from photospheric pressure modes to Alfvénic waves after $z\approx 2$~Mm. This result compliments the recent results of \citet{morton_2025}, who discuss high-resolution observations of Doppler-velocity fluctuations derived from DKIST Cryo-NIRSP data. In this study they suggest 3.5~mHz Alfvénic fluctuations originating as photospheric p-modes, driven by convection. We note that the second dominant frequency observed and discussed by \citet{morton_2025}, corresponding to $6.38$~mHz does not feature significantly in our MSF results. \\

In the NS region, oscillations of 5~mHz are dominant. In the vertical slice, Figure \ref{fig: MSF_vert}, and time-distance plot, Figure \ref{fig: t-d_upar_p}, we see that the regions of 5~mHz in $p$ and $u_\parallel$ coincide. Additionally, we see that these regions also coincide with regions dominated by the same frequency in the vertical magnetic field, $B_z$, magnitude of the magnetic field, $|\mathbf{B}|$, and velocity perpendicular to the magnetic field, $u_\perp$. In order to see oscillations in all of these variables, we speculate the mode to be a vertically-propagating magnetoacoustic sausage mode. We disregard the possibility of a purely rotational (i.e., torsional Alfvén) mode due to the oscillations in $|B|$, although we note that a rotational component to the mode may be possible alongside the sausage mode. The origin of these oscillations is unclear: we see a small reservoir of 5~mHz frequencies in the convection zone (not shown) and photosphere, which suggests the convection as an acoustic driver. It is also possible that the driver of these coronal oscillations is related to the motion of the jet motions, such as type II spicules. This is plausible for two regions: firstly, that these frequencies are enhanced in the quiet plasma, where the spicules is not suppressed by the complex magnetic structure and velocities of the swirl, and secondly, since we see that the location of the appearance of 5~mHz frequencies has a weak correlation to the peaks of the spicules. This speculation once again coincides with the discussion by \citet{morton_2025}, on the unknown origin of their 6.3~mHz oscillations. The origins of this mode therefore harbour a rich opportunity for future research.

\subsection{Effects of swirl structure} \label{sec: swirl_discuss}
Our comparison of the two regions of the vertical slice has led to some stark differences between oscillations in the absence of a swirl and those in the presence of such a structure, which are summarised in Figure \ref{fig: sketch}. First and foremost, we see the suppression of 5~mHz oscillations, and enhancement of wave power in 3.5~mHz oscillations. This result encompasses the entire lifecycle of the swirl, and as such it is important to consider that the emergence of the swirl may present different results temporarily. In this case, as the swirl emerges we observed a temporary suppression on the upwards propagation of 3.5~mHz frequencies at the base of the swirl in the photosphere; a disparity from the general trend of 3.5~mHz enhancement in these regions. In future work it would therefore be interesting to study the evolution of the swirl (see \cite{Finley_2022},\cite{tziotziou_2023}) with respect to the effects on frequency propagation in time.\\

The slice $V1$ is a unique opportunity to compare two clearly differing environments in a realistic simulation. Even within this simulation, it is challenging to find another slice with the same proportion of swirling and quiet plasma. Slice V2 appears to also contain a large portion of quiet plasma for some time in the domain, but the interaction of the smaller swirl S3 along this slice limits the comparison. In this case, the results show a strong presence of 3.5~mHz and less 5~mHz oscillations throughout the domain. It is important to consider that the swirls are not fixed in location, and therefore even the results of slice V1 contain a mixture of the centre and boundary interactions of the swirl. It is important for future studies to differentiate between the wave propagation in swirl-interaction regions, boundaries, and the centre of swirls. Therefore, a slice through the centre of a swirling structure would be an important future subject. However, we note this was not possible here on S1 without losing the crucial non-swirling comparison region for our study. Furthermore, the highly dynamic nature of these structures makes consistently capturing the centre of a swirl a challenging feat without a dynamic frame of reference. We suggest slices V2 and V3 as notable slices for future studies. The slice V2 contains the multiple eruptions of the larger swirl S2, alongside plasma which interacts between S3 and S2. V3 captures the short-lived evolution of S3 and S4, which would require a narrower time-frame to truly compare only the swirling mechanisms, but could provide useful insights into the interaction region between them compared to inside the swirl structure. Future studies may also consider using a 3D approach to fully clarify a swirls role in the upward propagation of wave modes. \\

\subsection{Energy transfer}
We see significant contributions (up to 35$\%$) to the total wave power (both acoustic and magnetic) in regions with 3.5~mHz and 5~mHz MSFs at different heights. We show that a larger coverage of MSFs does not always correlate to a larger wave power, and vice versa. This could mean that a less wide-spread MSF could potentially contain a significant fraction of the wave power, outside of those studied in this paper. Nevertheless, we show that MSFs at 3.5 and 5~mHz contribute to up to 50$\%$ of the total wave power combined. Furthermore, the magnetic wave power is sustained in the upper atmosphere, only decreasing marginally, supporting the successful propagation of energy up to and above 6~Mm. The dissipation of such energy has not been discussed here, although this is also possible to detect from the MSF decomposition (see \citetalias{Cherry_2025}). This is therefore considered a next step in order to fully quantify the role of wave heating in the corona.

\begin{acknowledgements}
This project has received funding from the European Union's Horizon 2020 research and innovation programme under the Marie Skłodowska-Curie [grant agreement Nº 945371]. It is also supported by the Research Council of Norway through its Centres of Excellence scheme, project number 262622, and through grants of computing time from the Programme for Supercomputing.

A.J.F and Q.N received funding from the European Research Council (ERC) under the European Union’s Horizon 2020 research and innovation programme (grant agreement No 810218 WHOLESUN).

\end{acknowledgements}

\bibliographystyle{aa}
\bibliography{Main}

\begin{appendix}
\section{Higher frequencies} \label{app: high_freq}
\begin{figure*}[h]
    \centering
    \includegraphics[width=\linewidth]{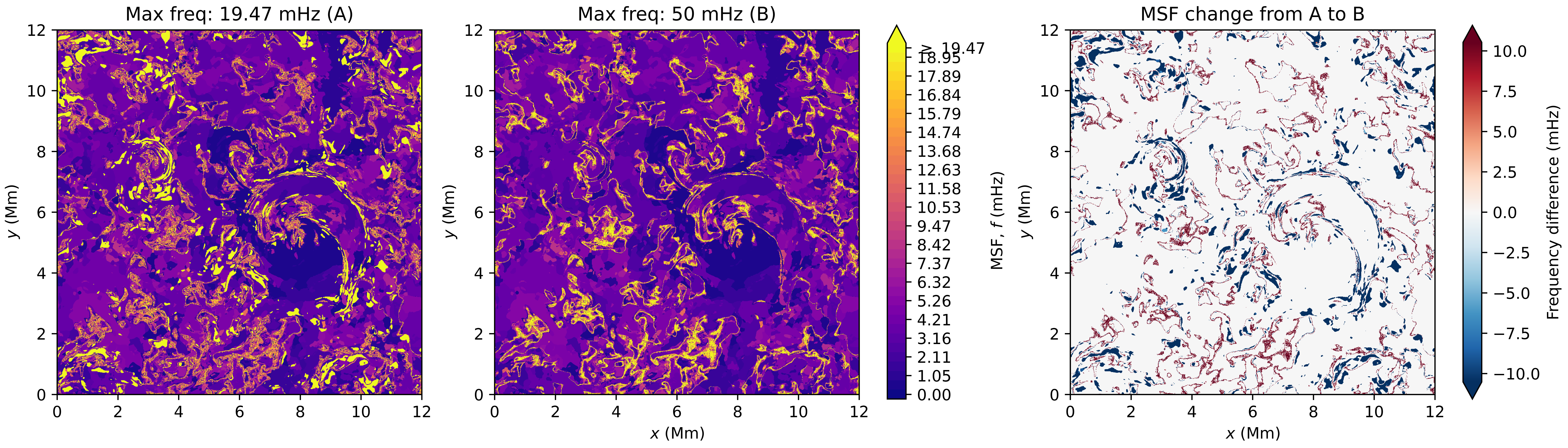}
    \caption{Comparison of two MSF decompositions on the velocity parallel to the magnetic field at $z=4$~Mm, time $t=1000$~s. The left-most panel shows the results for a maximum individual MSF of $19.47$~mHz considered in the decomposition, and is the same result shown in \citetalias{Cherry_2025}. In comparison, the middle panel shows the results for a maximum individual frequency of $50$~mHz considered in the decomposition. The change in MSFs between the two decompositions is displayed in the rightmost panel.}
    \label{fig: ext_msf}
\end{figure*}

The highest frequency (HF) bin in the MSF decomposition is an amalgamation of any and all frequency signals that are above the maximum frequency considered individually by the algorithm. This is necessary for computational efficiency, and, as shown in Section \ref{sec: shocks}, the numerical bi-products of this choice can be used to detect shocks if tuned correctly. Figure \ref{fig: ext_msf} shows the difference in results from calculating the MSFs with a maximum individual frequency of 19.47~mHz (upper panel, A) and 50~mHz (lower panel, B). \\

The points displayed in yellow for B now relate to the locations where the MSF is a single frequency above 19.47~mHz compared to A where the yellow points relate to the locations where the MSF is the combined signal of all frequencies higher than 19.47~mHz. This makes the high frequency results in B more refined. $50\%$ of the locations represented by the HF bin in A are replaced by lower frequencies in B, with an average decrease of 7.5~mHz (15 frequency steps). In contrast, only $\sim\%1$ of the lowest frequencies (>0~mHz) change between the two cases, with an average change of 1 frequency step (0.5~mHz). These changes are demonstrated in the bottom panel of Figure \ref{fig: ext_msf}. \\

By eye, it is possible to see that the "speckled" pattern of higher frequencies in A is largest replaced by the HF bin in B. This is due to the generally diminishing amplitudes of the higher frequencies in the DFT decomposition, which will match the lowest amplitudes of the original signal in the MSF calculation, thereby strengthening the role of "speckled" patterns for detecting damped oscillations (see \citetalias{Cherry_2025} for a full explanation of dissipation detection in the MSF decomposition).

\end{appendix}

\end{document}